\documentstyle[amssymb,preprint,prl,aps,12pt]{revtex}

\begin{document}

\begin{center}
{\Large ORIENTATIONAL ORDERING IN\ MONOLAYERS OF ORTHO-PARA HYDROGEN }

{\Large \ \bigskip }

{\Large \ V.B. Kokshenev}$^{1}$ and {\Large N.S. Sullivan}$^{2}${\Large \
\bigskip }
\end{center}

$^{1}$Departamento de F\'{i}sica, Universidade Federal de Minas Gerais,
Caixa Postal 702, 30123-970, Belo Horizonte, Minas Gerais, Brazil.

$^{2}$Physics Department, University of Florida, PO Box 118400, Gainesville,
FL 32611-8400.

.

(Submitted to the Fizika Nizkih Temperatur 27 of October, 2002 and in
revised form 6 of February, 2003).

\begin{center}
{\Large ABSTRACT}

.
\end{center}

We discuss orientational ordering in monolayers of solid hydrogen in view of
recent experimental findings in NMR studies of ({\it ortho)}$_{c}$-({\it %
para)}$_{1-c}$-hydrogen mixtures on boron nitride substrate. Analysis of the
temperature-concentration behavior for the observed NMR frequency splitting
is given on the basis of a two-dimension ($J=1)_{c}$-($J=0)_{1-c}$-rotor
model with the quadrupolar coupling constant $\Gamma _{0}=0.50\pm 0.03$ $K$
and the crystalline field amplitude $V_{0}=0.70\pm 0.10$ $K$ derived from
experiment. The two distinct para-rotational (PR) short-range ordered
structures are described in terms of the local alignment and orientation of
the polar principal axis, and are shown to be due to the interplay between
the positive and negative crystalline fields. It is shown that the observed
below the 2D site percolation threshold $c_{p}=0.72$ local structures are
rather different from the ferromagnetic-type PR ordering suggested earlier
by Harris and Berlinsky.


.

Corresponding author Valery B. Kokshenev, valery@fisica.ufmg.br

\newpage

\section{INTRODUCTION}

Careful studies at low temperatures of the thick films (from 2 to 12
monolayers)\cite{KimS98,KimS99} and monolayers\cite{KimS99,KimS97} of
ortho-para hydrogen on boron nitride (BN) substrates revealed at low ortho-$%
H_{2}$ concentrations new {\em short-range} frozen structures. Besides the
analogue of known\cite{SDC78,HM85,K85,LS87,KL87,HKL90,K91,BR92,W92,KS01} 
{\em quadrupolar glass} (QG) phase, that emerges\cite{KimS99} in monolayers
below the concentration $c_{p}=0.72$ , which is apparently close to the
site-percolation threshold\cite{Isi92} in honeycomb lattice, the uncommon 
{\em para}-{\em rotational} (PR) {\em phases} denominated by PR-A and PR-B
phases were discovered\cite{KimS99,KimS97}. They are demarcated by a
crossover temperature $T_{x}^{(\exp )}(c)$ at which the NMR frequency
splitting passes through zero (see 2D diagram in {\bf Fig.1}). Instead of
the $Pa3$ structure known in the bulk hydrogen, the herring bone (HB) and
pinwheel (PW) 2D long-range ordered structures have been the subject of
scientific interest since 1979, when Harris and Berlinsky made\cite{HB79}
their famous mean-field theory predictions. Meanwhile thorough experimental
studies on grafoil\cite{KH78,KHG85} and BN\cite{KimS99,KimS97} substrates
registered only the PW orientational order at sufficiently high
concentrations, {\it i.e.} above the percolation limit $c_{p}$.

Analysis is given within the scope of the site-disordered microscopic 2D $%
(J=1)_{c}(J=0)_{1-c}$-rotor model, which is introduced on the basis of the
3D-rotor analog developed earlier\cite{PRB96,JLTP96,JLTP2,JLTP98} for an
indepth study of the QG phase. We will give a microscopic explanation of the
observed temperature-concentration behavior for the orientational
local-order parameters related to the NMR line shapes. We will show that the
PR-A and PR-B short-range correlated structures are due to the interplay of
the frustrated $o$-$H_{2}$-molecular exchange interaction with the
molecule-substrate interaction.

\section{MICROSCOPIC DESCRIPTION}

Description of the orientational degrees of freedom of the site-disordered
ortho-para-hydrogen system with pure electrostatic quadrupole-quadrupole
(EQQ) intermolecular interactions has been discussed extensively within
context of the 3D QG problem\cite{K85,KL87,W92,PRB96}. In general, the
thermodynamic rotational states of a given ortho-molecule located at a site $%
i$ are characterized by the a second rank {\em local} tensor that has only
five independent components: the three principal local axes (given by vector 
${\bf L}_{i}$) , and the {\em alignment} $\sigma _{i}=<(1-3\stackrel{\wedge 
}{J}_{zi}^{2}/2)>_{T}$ and the {\em eccentricity} $\eta _{i}=<\stackrel{%
\wedge }{J}_{xi}^{2}-\stackrel{\wedge }{J}_{yi}^{2}>_{T}$ defined with
respect to ${\bf L}_{i}$ axes. (Here $\widehat{J}_{\alpha i}$ stands for the
angular-operator rotational moment of a given ortho-molecule located\cite
{KL87} at site $i$, and $<...>_{T}$ refers to a {\em thermodynamic average}
at temperature $T$.) A thermodynamic description, given in terms of the {\em %
local molecular fields} ${\varepsilon _{\sigma i}}$, and $\varepsilon _{\eta
i}$ conjugate to the local order parameters and extended by the {\em %
crystalline field} $h_{i}$ can be introduced\cite{PRB96,JLTP96} on the basis
of the local-order-parameter fundamental equations, namely

\begin{eqnarray}
\sigma _{i} &=&1-\frac{3\cosh (\sqrt{3}\varepsilon _{\eta i}/2T)}{2\cosh (%
\sqrt{3}\varepsilon _{\eta i}/2T)+\exp [3{(\varepsilon _{\sigma i}+}h_{i}{)}%
/2T]};  \eqnum{1a} \\
\eta _{i} &=&\frac{\sqrt{3}\sinh (\sqrt{3}\varepsilon _{\eta i}/2T)}{2\cosh (%
\sqrt{3}\varepsilon _{\eta i}/2T)+\exp [3{(}\varepsilon _{\sigma i}{+}h_{i}{)%
}/2T]},  \eqnum{1b}
\end{eqnarray}
These equations follow from the conditions of local equilibrium\cite{K91}
and they are shown\cite{JLTP96} to be consistent with the density-matrix
representation\cite{LS87}. In what follows, we restrict our consideration to
a reduced set of local order parameters $\{{\bf L}_{i},\sigma _{i}\}$ with $%
\eta _{i}\equiv 0$ that corresponds to the so-called ''powder
approximation'' common\cite{HM85} in NMR theory applications. This
description ignores the local field $\varepsilon _{\eta i}$, conjugate to
the local eccentricity $\eta _{i}$, and Eq.\thinspace (1) (where $%
\varepsilon _{\sigma i}=\varepsilon _{i}$ and $\varepsilon _{\eta i}=0$) is
therefore reduced to 
\begin{equation}
\frac{1-\sigma _{i}}{1+2\sigma _{i}}=\exp [{-}\frac{3}{2}\frac{{(}%
\varepsilon _{i}+h_{i}{)}}{T}]\text{ with }{\varepsilon _{i}=-}\sum_{j\neq
i}^{z}{\cal {J}}_{ij}\sigma _{j}c_{j}\text{.}  \eqnum{2a}
\end{equation}
Here the effective {\em exchange interaction} ${\cal {J}}_{ij}$ and the
crystalline field $h_{i}$ are given by 
\begin{equation}
{\cal {J}}_{ij}=-\frac{3}{2}\Gamma _{0}P_{20}(L_{zi})P_{20}(L_{zj})\ \ \ 
\text{and}{\rm \ \ \ }h_{i}=\frac{2}{3}V_{0}P_{20}(L_{zi}).  \eqnum{2b}
\end{equation}
This reduced mean-field description formally follows from the truncated 2D
Hamiltonian given for $N$ quantum rotors with $z$ neighbors placed in the
plane, namely

\begin{equation}
\stackrel{\wedge }{H}_{N}=-\sum_{i=1}^{N}\sum_{j\neq i}^{z}{\cal {J}}_{ij}%
\stackrel{\wedge }{\sigma }_{i}\widehat{\sigma }_{j}c_{i}c_{j}-%
\sum_{i=1}^{N}h_{i}\stackrel{\wedge }{\sigma }_{i}c_{i}\ \text{with}%
\stackrel{\wedge }{\sigma }_{i}=1-\frac{3}{2}\stackrel{\wedge }{J}_{zi}^{2}.
\eqnum{2c}
\end{equation}
Here $c_{i}$ is a random occupation number whose mean, given by the {\em %
configurational average}, is the concentration: $c=<c_{i}>_{_{C}}$; $%
\stackrel{\wedge }{J}_{zi}$ is a $z$-projection of the angular momentum
operator in the local principal coordinate system. In turn, $\Gamma _{0}$
stands for the EQQ{\em \ coupling constant} and $V_{0}$ is the {\em %
crystal-field amplitude}; $P_{20}(L_{zi})=(3\cos ^{2}\Theta _{i}-1)/2$ where 
$\Theta _{i}$ is the polar angle of the principal molecular axis ${\bf L}_{j}
$. 

In bulk HCP solid ortho-para-hydrogen the coupling constant $\Gamma _{0}$
and the amplitude $V_{0}$ are well established theoretically that is not the
case of commensurate $\sqrt{3}\times \sqrt{3}$ solid monolayers. The
magnitudes of $\Gamma _{0}={6Q^{2}}/{{(25R_{0}^{5})}}$ ($Q$ is the molecular
electrostatic quadrupole moment and $R_{0}$ is the nearest-neighbor
molecular separation) calculated for graphite and BN substrates are, $0.534$ 
$K$\cite{HB79} and $0.470$ $K$\cite{KimS99}, respectively. The approximate
estimates for the crystal-field amplitude $|V_{0}^{(\exp )}|=0.6-0.8$ $K$
for grafoil\cite{Kim-Diss} and $|V_{0}^{(\exp )}|\approx 0.6$ $K$ for BN\cite
{KHG85} were derived from the observed NMR line shapes. As seen from Fig.1,
the experimentally established position of the order-disorder boundary is
consistent for the cases of grafoil\cite{KH78} and BN\cite{KimS99}
substrates (shown by crosses and triangles in Fig.1, respectively). On the
other hand, the PW-PR boundary (restricted by points $a$ and $b$ in the
insert of Fig.1) exists only for {\em positive }crystalline fields. These
yield the following fundamental model parameters

\begin{equation}
\Gamma _{0}=0.50\pm 0.03K\text{ and }V_{0}=0.70\pm 0.10K  \eqnum{2d}
\end{equation}
in order to specify our estimations based on the 2D $(J=1)_{c}(J=0)_{1-c}$%
-rotor model given in Eq.(2).

\section{Macroscopic Description}

The phase diagram for the pure $J=1$ rotor system on a 2D triangle solid
lattice was scaled\cite{HB79} by Harris and Berlinsky in terms of the EQQ
coupling constant $\Gamma _{0}$ and the crystal-field amplitude $V_{0}$ of
both signs (see insert in Fig.1). The PR phase was postulated by a single
ferromagnetic-type structure that can be given as $\{\Theta _{i}=0,\sigma
_{i}=\sigma _{0}\}$. Moreover, one can see that equation for the PR
alignment $\sigma _{0}$ obtained by minimization of the relevant free energy
(see Eq.(17) in Ref.\cite{HB79}) is equivalent to Eq.(2) with adopted $z=6$, 
$\varepsilon _{i}=\varepsilon _{0}=-9\Gamma _{0}\sigma _{0}$, ${\cal {J}}%
_{ij}=-3\Gamma _{0}/2$ and $h_{i}=h_{0}=2V_{0}/3$ that corresponds to a
ferrorotational-type (FR-type) local-structure given by $\Theta _{i}=\Theta
_{j}=0$ and $\sigma _{i}=\sigma _{0}$.

A description for the long-range orientationally disordered, but locally
correlated PR-A phase is introduced through the {\em short-range} order
parameter $\sigma _{A}(c,T)=<\sigma _{i}(T)>_{_{C}}^{(PRA)}$, where a
configurational average is limited by the temperature-concentration PR-A
region shown in Fig.1. Application of this average procedure to both the
sides of Eq.(4) can be presented in the following form, namely 
\begin{equation}
\text{PR-A:}\;\frac{1-\sigma _{A}}{1+2\sigma _{A}}={\exp }\left[ -\frac{V_{0}%
}{T}-\frac{3}{2}(\frac{\varepsilon _{1A}+\delta \varepsilon _{1A}}{T})+\frac{%
9}{8}(\frac{\varepsilon _{2A}}{T})^{2}\right] .  \eqnum{3a}
\end{equation}
Unlike the case of the QG, we assume here that fluctuations of the local
alignment (or the {\em quadrupolarization}) are small. The same is referred
to the crystalline field given by the mean $h_{1A}=2V_{0}/3$. The local
fluctuations of the molecular field are introduced through the mean $%
\varepsilon _{1A}={\cal {J}}_{1A}zc^{5/2}\sigma _{A}$ (with ${\cal {J}}%
_{1A}=-3\gamma _{1A}\Gamma _{0}/2$ and with $z=6$) and the variance $%
\varepsilon _{2A}^{2}=3{\cal {J}}_{2A}^{4}zc(1-c)(1-\sigma
_{A}^{2})^{2}/8T^{2}$, and are estimated\cite{gauss} within the Gaussian
distribution justified in Ref.\cite{JLTP96}). With account of the
Zeeman-field local polarization effects\cite{PRB96} given by the mean $%
\delta \varepsilon _{1A}=-{\cal {J}}_{2A}^{2}zc\sigma _{A}(1-\sigma
_{A}^{2})/T$, we have analyzed\cite{est} a concentration behavior of the
observed\cite{Kim-Diss} NMR frequency splitting given in Fig.\thinspace 8 in
Ref.\cite{KimS99} for $T=0.65$ $K$ and $T=0.546$ $K$. Analysis is given with
the help of Eq.(2) where the polar-principal-axis correlation parameters $%
\gamma _{1A}$ and $\gamma _{2A}$, namely

\begin{equation}
\gamma _{1A}=<P_{20}(L_{zi})P_{20}(L_{zj})>_{C}^{(PRA)}\text{and }\gamma
_{2A}=\sqrt{<P_{20}^{2}(L_{zi})P_{20}^{2}(L_{zj})>_{C}^{(PRA)}}  \eqnum{3b}
\end{equation}
are treated as fitting parameters. In the particular cases of the FR-type ($%
\Theta _{i}=\Theta _{j}=0$) and AFR-type ($\Theta _{i}=0,\Theta _{j}=\pi /2$%
) locally correlated structures are characterized by $\gamma _{1}=\gamma
_{2}=1$ and $\gamma _{1}=-\gamma _{2}=-1/2$, respectively. For the PR-A
phase we have derived\cite{tobe} $\gamma _{1A}^{(\exp )}=-{1/3}$ and $\gamma
_{2A}^{(\exp )}\thickapprox 0.75$. This local structure is in a way similar
to that in the PW phase modified by orientations of in-plane rotors which
show out-of-plane orientations.

As seen from Fig.1, the long-range disordered PR-B phase is stable at low
temperatures ($T<V_{0}$) and low concentrations ($c<c_{p}$) where the
site-dilution effects are expected to be more pronounced than in the PR-A
phase. The short-range orientational arrangement results from the interplay
between the random EQQ coupling and the random {\em negative}\ crystalline
fields. Adopting for the latter a Gaussian distribution, and taking into
account its variance $h_{2B}$ (with the mean $h_{1B}=2V_{0}/3$) one finds,
after elaboration\cite{gauss} of the configurational average in Eq.(2a), the
effective amplitude of the crystalline field can be introduced as

\begin{equation}
V(c,T)=V_{0}(1-\frac{T_{x}{^{(\exp )}(c)}}{T})\text{ for }T{\thickapprox }%
T_{x}{^{(\exp )}(c)}\text{.}  \eqnum{4}
\end{equation}
Here $T_{x}{^{(\exp )}(c)}$ is a crossover temperature between the PR-A and
the PR-B structures that provides a reconstruction of the local order from $%
\sigma _{A}^{(\exp )}(c,T)>0$ to $\sigma _{B}^{(\exp )}(c,T)<0$ (shown by
dashed line in Fig.1). The explicit form in Eq.(4) follows from $%
h_{2B}=(<\Delta h_{i}^{2}>_{_{C}}^{(PRB)})^{1/2}=2(2V_{0}T_{x})^{1/2}/3$
where $T_{x}$ is approximated by the observed PRA-PRB boundary. An analysis
of the observed PR-B quadrupolarization $\sigma _{B}$ is given through
averaged Eq.(4), namely 
\begin{equation}
\text{PR-B:}\ln (\frac{1-\sigma _{B}}{1+2\sigma _{B}})+\frac{V_{0}}{T}(1-%
\frac{T_{x}{^{(\exp )}(c)}}{T})-9(\frac{\gamma _{2B}\Gamma _{0}}{T}%
)^{2}c\sigma _{B}(1-\sigma _{B}^{2})=0.  \eqnum{5}
\end{equation}
Treating the PR-B phase as a precursor of the 2D QG phase, we have omitted
in Eq.(5) all molecular-field local ordering effects. Similar to the QG
case, we have therefore adopted $<\varepsilon _{i}>_{C}^{(PRB)}=\delta
\varepsilon _{1B}$ employed in Eq.(3a) for the PR-A phase. Analysis of the
available experimental data for $c=0.44$ (with $T_{x}^{(\exp )}=1.64$ $K$ ,
see Fig.\thinspace 12 in Ref.\thinspace \cite{KimS99}) on temperature
dependence of the short-range orientational order parameter in the PR-B
given with the help of Eq.(5) results\cite{tobe} in, approximately, $\gamma
_{1B}=0$, $\gamma _{2B}=1$, that in a way is characteristic for the QG local
order.

The observed order parameters $\sigma _{A}^{(\exp )}$ and $\sigma
_{B}^{(\exp )}$ vanish at a certain crossover temperature $T_{x}(c)$
associated with the PRA-PRB boundary $T_{x}^{(\exp )}(c)$ (shown by the
dashed line Fig.\thinspace 1). For concentrations $c\leq c_{p}$, this
boundary can be therefore deduced from the conditions $\sigma
_{A}(c,T_{x})=\sigma _{B}(c,T_{x})=0$. To satisfy the boundary observation
conditions, the interplay between the fluctuating crystalline and
Zeeman-type molecular fields for $T{_{0}<}T\leq T_{x}$ is made implicit in
the form $8{V_{0}(}T-T{_{0}})-9\varepsilon _{2}^{2}(c,T)=0$, where $T{_{0}}$
($=9h_{2}^{2}/8V_{0}$) plays a role of $T_{x}$ when the competing
fluctuations the EQQ field are ignored. The variance $h_{2}$ was studied\cite
{JLTP2} in detail for the 3D disordered PR phase in {\it o-p}-$H_{2}$
systems for temperatures $0.80<T<4.9$ $K$ . As seen from the right insert in 
{\bf Fig.2}, unlike the mean of the crystal-field $h_{1}^{(3D)}$, its
variance depends strongly on the overall concentration, {\it i.e}. , $%
h_{2}^{(3D)}\sim c(c_{M}-c)$ and disappears at the highest concentration for
the 3D QG state, $c_{M}=0.55$ (for the 3D phase diagram see Fig.\thinspace 2
in Ref.\cite{JLTP96}\thinspace ). In the 2D case $c_{M}$ is very close to
the threshold concentration $c_{p}$. Therefore, we adopt $V_{2}$ $\sim
c(c_{p}-c)$ that reduces the aforegiven boundary observation condition to
the following cubic equation 
\begin{equation}
T_{x}^{3}-T_{0}(c)T_{x}^{2}-\left( \frac{3}{2}\gamma _{2}\Gamma _{0}\right)
^{4}\frac{c(1-c)}{2V_{0}}=0{\rm \ \ \ }\text{with}{\rm \ \ \ }%
T_{0}=8V_{0}\left[ \lambda \frac{c}{c_{p}}(1-\frac{c}{c_{p}})\right] ^{2}. 
\eqnum{6}
\end{equation}
Treating $\lambda $ as an adjustable parameter characterizing a scale of the
crystal-field fluctuations, we analyze in Fig.\thinspace 2 the physical
solution $T_{x}$ of Eq.\thinspace (6) by comparing it with the observed
PRA-PRB boundary. Taking into account the above analysis for PR-B and PR-A
phases, we adopt $\gamma _{2}=1$ as a typical value. As seen from Fig.2, the
idea that the disordered PRB phase is constructed from mostly disordered
''in-plane'' rotors is corroborated by experimental observations\cite{KimS99}%
. On the other hand, our consideration of the reduced orientational degrees
of freedom fails to give quantitative descriptions above $c=0.45$. The
unphysical value $\gamma _{2}>1$ deduced from experimental data in Fig.2
signals the existence of ignored local order parameters ({\it e.g. }%
\thinspace $q_{\sigma }\neq \sigma ^{2}$), which (similar to the case of the
3D QG given in Eq.(6) in Ref.\cite{PRB96}) can play an appreciable role near
the crossover temperature. A complete analysis should be given beyond the
''powder approximation'' and based on fundamental order-parameter equation
(1) given for both the local alignment and the eccentricity.

\section{CONCLUSIONS}

We have discussed the short-range orientationally correlated structures
discovered\cite{KimS99,KimS97} in monolayers of ($o$-$H_{2}$)$_{c}$($p$-$%
H_{2}$)$_{1-c}$ on BN substrate. Analysis of the temperature-concentration
behavior for the observed NMR line shapes, related to the short-range order
parameter $\sigma (c,T)$, is given on the basis of the 2D ($J=1)_{c}$-($%
J=0)_{1-c}$-rotor model, for which a 3D analog was employed earlier\cite
{PRB96,JLTP96} for the QG problem. In the current study the focus is on the
nearest-neighbor correlated structures observed by the NMR spectroscopy in
the orientationally disordered phases (shown in Fig.1). In spite of the fact
that fundamental order-parameter equations are consistent with the
corresponding Eq.(17) in Ref.\cite{HB79}, the observed PR-A and the PR-B
structures are rather ''antiferromagnetic'' than ''ferromagnetic'' as
suggested in Ref.\cite{HB79} for a unique PR phase. This conclusion follows
from our analysis of the observed\cite{KimS97,KimS99} macroscopic
quadrupolarizations $\sigma _{A}(c,T)$ and $\sigma _{B}(c,T)$ adjusted
through the polar-axis correlation parameters given in Eq.(3b). We have
shown that the short-range correlated PR-A phase is driven by positive
crystalline fields, for which thermal and spatial fluctuations overwhelm
those of the short-range EQQ interactions. With decreasing temperature, the
interplay between the EQQ coupling and the crystalline fields, which are
both sensitive to the site-dilution and thermal-fluctuation effects, results
in the PRA-PRB\ boundary $T_{x}^{(\exp )}(c)$, along which both the
quadrupolarizations are zero (see analysis in Fig.2). The low-temperature PR
phase, denominated as the PR-B phase, is driven by {\em negative }%
crystalline field given near the boundary $T_{x}^{(\exp )}$ in Eq.(4).
Similar to the case of the QG phase (see Fig.4 in Ref.\cite{JLTP96}), this
phase is expected to be richer than the PR-A phase, and more order
parameters are therefore needed to give a complete description of the
observed $\sigma _{B}^{(\exp )}(c,T)$. Unfortunately, this data (such as on $%
\Delta q_{B}=<$ $\sigma _{i}^{2}(T)-\eta _{i}^{2}(T)>_{_{C}}^{(PRB)}$) is at
present time not available experimentally.

Acknowledgments - The authors acknowledge financial support by the CNPq
(V.B.K.) and by the NSF-DMR-98 (N.S.S.).

{\Large Figure Captures}

Fig.1. Phase diagram for site-disordered monolayers of ortho-para-hydrogen $%
(o$-$H_{2})_{c}(p$-$H_{2})_{1-c}$ mixtures. The symbols refer to observed
changes in the NMR line shapes reported in the literature: {\em crosses} for
hydrogen monolayers on graphite, Ref.\thinspace \cite{KH78}; {\em open
triangles} for commensurate hydrogen monolayers on BN, Ref.\thinspace \cite
{KimS99}. The solid symbols refer to NMR studies of Ref.\cite{KimS99}: {\em %
solid circles}, transitions to the quadrupolar glass (QG) state; {\em %
diamonds}, transitions to the hindered rotor (HR) state. The {\em inverted
triangles} refer to the vanishing of the small splitting of the NMR lines in
the para-rotational state. {\em Insert}: theoretical phase diagram from
Fig.\thinspace 2 in Ref.\cite{HB79}; $a$ and $b$ are the tricritical points%
\cite{HB79}, and $c$ is the minimum in the observed PR-PW transition
temperatures.

\vskip0.12in

\noindent Fig.2. Pararotational A-B crossover temperature against
concentration. The symbols refer to the experimental points represented in
Fig.\thinspace 1: {\em solid inverted triangles}, vanishing of NMR doublet,
Ref.\cite{KimS99}; {\em open triangles}, onset of PW state in the anomalous
upturn region of the phase transition boundary, Ref.\cite{KimS99}. The {\it %
lines} designate solutions of Eq.\thinspace (6) for an adjustable parameter $%
\lambda =2.3$. Other parameters shown include the fitting parameter $\gamma
_{2}$. {\em Insert}: Left; Concentration dependence of the effective
crystalline field at distinct temperatures derived from experiment\cite
{KimS99} using Eq.(4); right, the variance of the crystalline field in bulk 
{\it ortho-para-}$H_{2}$ ({\em squares}, from insert in Fig.\thinspace 5 of
Ref.\thinspace \cite{JLTP2}).


\begin{references}
\bibitem{est}  The observable quadrupolarization is introduced by the
relation $|\sigma ^{(\exp )}(c,T)|=\nu (c,T)/3d$ where $\nu $ is the NMR
frequancy splitting and $d=57.67$ kHz.
\bibitem{gauss}  One can show that for the Gaussian average $<\exp (\pm
ax)>_{G}=$ $\exp [\pm ax_{1}-(ax_{2})^{2}/2]$ is true for a random value $x$
with mean $x_{1}$ and variance $x_{2}$.

\newpage
\end{references}
\end{document}